\documentclass%
[12pt]{article}

\usepackage{natbib}
\usepackage{amsfonts}
\usepackage{graphicx}
\usepackage{amssymb}
\usepackage{amsmath}
\usepackage[T1]{fontenc}
\newcommand{\scrif}{{{\mathcal I}^{+}}}

\title{Black hole uniqueness and magnetic shear}

\author{Adam D. Helfer\\%
Department of Mathematics and Department of Physics \& Astronomy\\%
University of Missouri\\%
Columbia, MO 65211, U.S.A.\\%
helfera@missouri.edu}
\date{\today}

\begin{document}

\maketitle
%% 125 w max

\begin{abstract}
A series of tantalizing results led to the black hole uniqueness conjecture:  isolated, realistic black holes should settle down to states characterized by their spin, mass and charge.  I argue that generically real black holes will also possess a `magnetic' shear; equivalently, that the dominant contribution to their long-range gravitational field should have a `magnetic' (odd-parity) component.  In fact, the Blandford--Znajek process, combined with the axial anomaly and gravitational gyrotropy, would tend to leave a black hole in such a state.  It seems that the black-hole uniqueness conjecture may apply in a regime around the hole, but as one approaches future null infinity
the `magnetic' effects become significant.  In this far-field regime, the space--time will be non-stationary, but there will be no radiation.
\end{abstract}

\bigskip\bigskip\bigskip
\centerline{Essay written for the Gravity Research Foundation 2018 Awards for Essays on Gravitation.}

\eject

%% 1500 w max

\section{Introduction}
\label{sec:intro}

Black holes are believed to be the end-states of general-relativistic gravitational collapse.  They relate vastly different scales; while they are defined by questions of when signals can escape `to infinity', they are compact objects.  There is a broad view, supported by a deal of circumstantial evidence, that real isolated astrophysical black holes ought to `settle down' to be those in the Kerr--Newman family, being characterized by their mass, spin and charge:  this is the black-hole uniqueness conjecture.\footnote{If non-electromagnetic contributions to the stress--energy are allowed outside the horizon, counterexamples can be constructed \citep{CCH2012}.}

I am going to describe a difficulty, not previously considered, contradicting this view:
there are strong reasons to think that real black holes will have a (perhaps small, but finite) persistent magnetic Bondi shear.  
This quantity is a spin-weight two function defined on the asymptotic sphere of directions; equivalently, it is the information in the $j\geq 2$ components of the Newman--Penrose curvature quantity $\Im\Psi _2^0$ \citep{PR1986}.  It therefore contains functional degrees of freedom, apparently strongly violating of the uniqueness conjecture.  But I will also
suggest a resolution of this conflict,
by examining with more care what it means for a black hole to `settle down'.

While the possibility of magnetic Bondi shear is evident from the asymptotic Newman--Penrose equations \citep[p. 395]{PR1986}, it has not received much attention.
Generating such shears requires a sort of odd-parity counterpart of a gravitational wave memory effect, which does not arise in simple explicit models.  It is also likely that the black-hole uniqueness conjecture has itself suggested to some that magnetic shear is somehow precluded.  
\cite{Winicour2014} argued that it might be forbidden in linearized gravity.
But in nonlinear general relativity, with no symmetry principle forbidding it, 
it seems clear that in generic situations magnetic shear must be present.

While I consider the foregoing argument convincing, I will also outline a specific mechanism which can generate magnetic shear for known astrophysical black holes.   This arises from a combination of three interesting processes:  (a) that of
 \cite{BZ1977}, which depends on a region in which the chiral electromagnetic invariant ${\bf E}\cdot{\bf B}\not=0$; (b) the axial anomaly, which means that in this region there will be a net current $J^5_a\not= 0$ of right- minus left-handed fermions \citep{Adler1969,BJ1969}; and (c) gravitational gyrotropy \citep*{ADH2016FGG}, which shows that gravitational waves created by matter falling into the hole, which would give `memory' effects, generate persistent magnetic Bondi shear after passing through the $J^5_a\not=0$ regime.

Quantitative estimates of the effect would be involved, and will not be attempted here.
It is likely the contribution of any one clump of infalling matter is very tiny for known holes, because the axial anomaly and the gravitational gyrotropy involve small prefactors, and also the ${\bf E}\cdot {\bf B}\not= 0$ region may not be favorably positioned.  But the effect may accumulate over the lifetime of the hole (the sign of the main contributions would not be expected to change unless the magnetic field around the hole reversed relative to the spin axis).

Even if the net effect of processes (a)--(c) is minute, that it may be non-zero carries broad significance.  It shows that there is no `hidden symmetry' forcing the magnetic shear to vanish.  We should expect these shears generically.

\section{Bondi--Sachs asymptotics}

In the Bondi--Sachs formalism, as refined by Newman and Penrose \citep{PR1986}, the Bondi shear $\sigma (u,\theta ,\phi )$ at future null infinity $\scrif$ plays a key role.  It is a spin-weight two quantity, and it can be written as $\sigma =\eth ^2\alpha$ for a spin-weight zero scalar $\alpha$.  The `electric' and `magnetic' parts of the shear --- really, they are the even and odd parity parts --- are $\eth ^2\Re\alpha$ and $i\eth ^2\Im\alpha$.
It is the $u$-derivative $\dot\sigma$ of the shear which codes gravitational radiation.

The electric part of the shear is closely tied to supertranslations.  Indeed, under a supertranslation $u\to u+\gamma (\theta ,\phi )$, we have $\sigma\to\sigma -\eth ^2\gamma$.
One sees immediately that in the absence of radiation, one can always find a supertranslation eliminating the electric part of the shear.  
It may happen that in two intervals of retarded time $u$ the shear is purely gauge (purely electric and $u$-independent), but nevertheless the shear in one is supertranslated with respect to the other.  This is the cleanest example of a {\em memory effect}, going back to \cite{BVM}.

The magnetic part is unaffected by supertranslations.  Indeed, the magnetic shear codes (when $\dot\sigma =0$) the same information as the Newman--Penrose curvature component $\Im\Psi _2^0$; it is gauge-covariant.  It is interpretable as spin (per unit energy) associated with the radiative degrees of freedom \citep{ADH2007}.  
A net change in magnetic shear from one $u$-independent value to another
has sometimes been called `magnetic memory', but it  
is not a memory effect in a conventional sense.
(Compare \cite{Winicour2014}.)

The asymptotic Newman--Penrose equations admit solutions where the system has settled down to a state where no radiation is being emitted ($\dot\sigma =0$) but there is a persistent magnetic shear.  
These will be our interest.
Such solutions cannot be stationary in the asymptotic regime; the equations imply that the curvature component $\Psi _1^0$ grows linearly, and $\Psi _0^0$ quadratically, with $u$.  However this statement depends strongly on nonlocal properties of the Bondi--Sachs gauge.  An asymptotic observer close to any one generator of $\scrif$ would not have access to the information needed to fix the Bondi--Sachs frame and verify this growth; she would in general be able to find a local frame in which the components did not grow.

\section{Consequences of the Blandford--Znajek scenario}

Many black holes are thought to be rotating and surrounded by accretion disks which bear magnetic fields.  Blandford and Znajek described an effect extracting energy from such holes.  There is a plasma which in the vicinity of the hole is `force-free', and this implies the electric ${\bf E}$ and magnetic ${\bf B}$ fields satisfy ${\bf E}\cdot {\bf B} =0$.  However, this equality breaks down far enough from the hole, and the break-down leads to electron--positron pair creation.
What will be most important for us, however, is that ${\bf E}\cdot {\bf B}\not=0$ will also lead to an imbalance $J^5_a\not= 0$ of left- and right-handed fermions, via the axial anomaly.   

Now suppose that (independent of those considerations) a bit of the accretion disk is drawn into the hole.  
The hole will be perturbed.  Near it, the metric will be complicated, but as one moves outwards to the wave zone the disturbance will resolve into radiation.  
If the infalling matter starts from
many gravitational radii from the hole, to leading approximation
it appears to come from infinity, and this means there will be a net change in the second  time derivative of the quadrupole moment.  We would therefore expect the outgoing gravitational wave to contribute a net `memory effect', or change in the electric part of the Bondi shear.

However, when a gravitational wave passes through a current $J^5_a\not= 0$, the effect is to rotate the polarizations of the right- and left-handed wave components in opposite senses \citep{ADH2016FGG}.\footnote{That paper emphasized the high-frequency limit, but its eqs. (A13), (A17), (A20) show the same result holds in our case (and its $\delta\psi\simeq 0$ in the radiation zone).}
In particular, the passage will interconvert fractions of the electric and magnetic shears.
Therefore the gravitational wave, as it passes outwards through the $J^5_a\not= 0$ region, will acquire a nontrivial magnetic Bondi shear.  Part of the jump in electric shear will be converted to a jump in magnetic shear.  While the electric part of the jump represents a memory effect, the magnetic one is a persistent change in the curvature component $\Im\Psi _2^0$.

\begin{figure}[h!]
  \includegraphics[width=\linewidth]{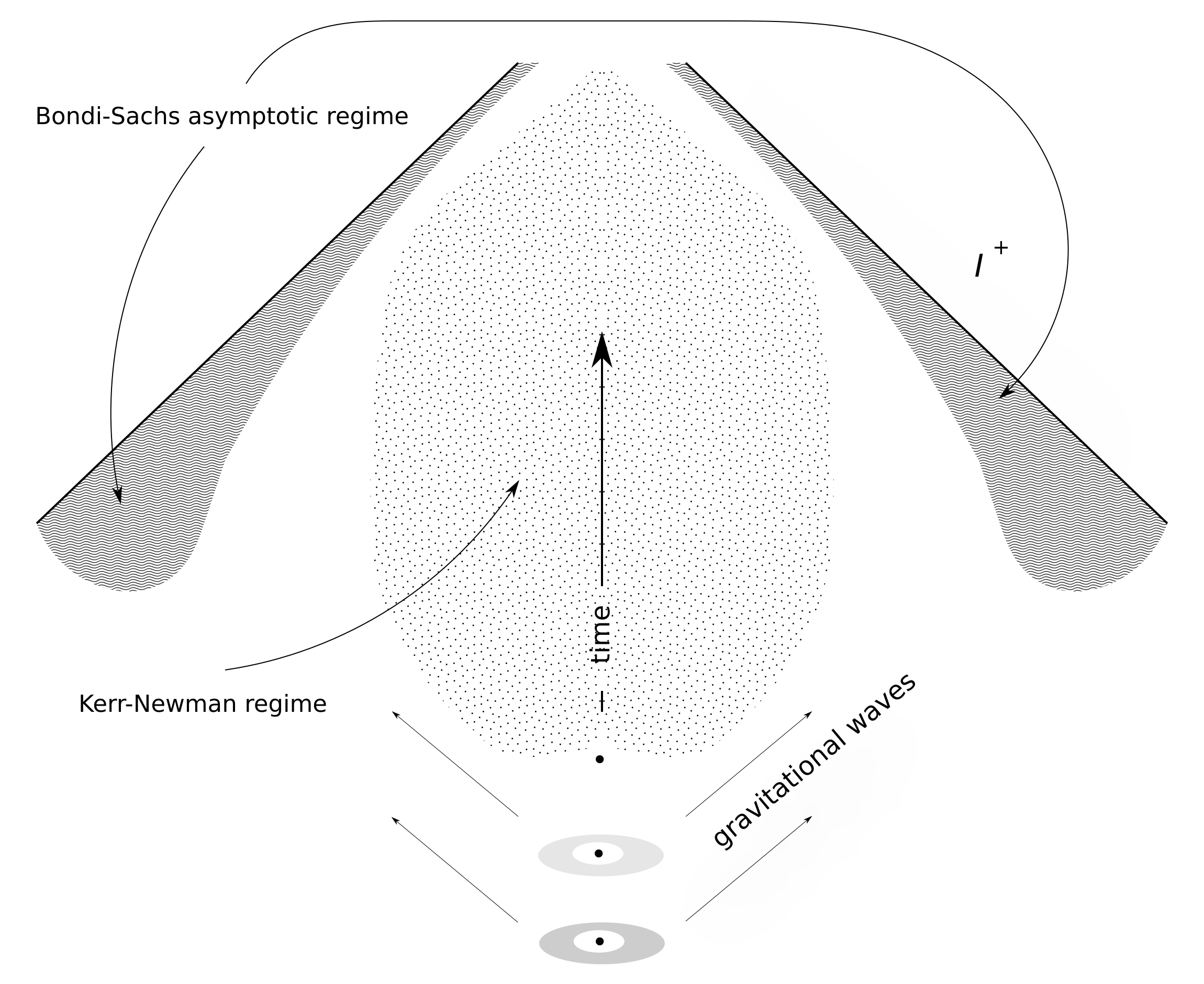}
  \caption{A schematic of the geometry described in the paper.  A black hole is surrounded by an accretion disk (bottom).  The disk is eventually exhausted, and gravitational radiation ceases.  In a growing region of space--time (stippled), the geometry approaches a Kerr--Newman solution.  However, moving outwards at the speed of light we pass to a region (shaded) in which the Bondi--Sachs asymptotics apply, a $u$-independent magnetic shear persists, and the space--time is not stationary.}
  \label{fig:stdiagram}
\end{figure}

\section{Reconciling the pictures}

We see there is good reason to expect that many black holes acquire, while matter accretes, magnetic Bondi shears.  These shears will persist unless they are countervailed by subsequent opposing odd-parity effects.   In particular, they should still be found once the discs are exhausted and the holes appear locally quiescent.

This contrasts sharply from the usual view that the hole should settle down to Kerr--Newman.  However, I do not think the views are really incompatible.  It is rather a question of different portions of the space--time `settling down' differently.  Most of the work on black-hole uniqueness does not speak to this dynamic process, but {\em assumes} that the relevant portion of the space--time has already become {\em stationary}, and attempts to classify the resulting possibilities. It is the fact that space--times with $u$-independent magnetic shear must fail (albeit somewhat subtly) to be stationary which opens a gap between that classification and the cases here.

It is possible that the region of space--time near the hole settles down and throws off all its $j\geq 2$ multipoles and so becomes close to Kerr--Newman, and also that this regime expands as time goes on, but that still, traveling outwards from there at the speed of light, one eventually exits that regime and makes a transition to one where the Bondi--Sachs asymptotics apply and the magnetic shear can be detected.  The inner regime would be the domain where black-hole uniqueness would apply, but the transition zone would ensheath this and beyond that would be the Bondi--Sachs regime.  See Figure~1.

The arguments here have been qualitative; it should be clear that quantitative analyses are now required.
It is possible that the effects predicted here --- $\Im\Psi _2^0\not=0$, linear evolution of $\Psi _1^0$ and quadratic of $\Psi _0^0$ --- could be detected by the Event Horizon Telescope (https://eventhorizontelescope.org).

\bibliography{ReferencesZ.bbl}

\bibliographystyle{agsm}

\end{document}